# On the thermal impact on the excavation damaged zone around deep radioactive waste disposal

P. Delage
*Ecole des ponts ParisTech, Navier/CERMES*

**Abstract**: Clays and claystones are considered in some countries (including Belgium, France and Switzerland) as a potential host rock for high activity long lived radioactive waste disposal at great depth. One of the aspects to deal with in performance assessment is related to the effects on the host rock of the temperature elevation due to the placement of exothermic wastes. The potential effects of the thermal impact on the excavated damaged zone in the close field are another important issue that was the goal of the TIMODAZ European research project. In this paper, some principles of waste disposal in clayey host rocks at great depth are first presented and a series of experimental investigations carried out on specific equipment specially developed to face the problem are presented. Both drained and undrained tests have been developed to investigate the drained thermal volume changes of clays and claystone and the thermal pressurization occurring around the galleries. This importance of proper initial saturation (under in-situ stresses) and of satisfactory drainage conditions (in spite of the significantly low permeability of claystones) is emphasized, leading to the development of a new hollow cylinder apparatus. It is observed that claystones cannot be considered as overconsolidated clays given that they can exhibit, as the Callovo-Oxfordian claystone does, a thermoplastic contraction. Mechanical and thermal hardening are however observed, extending to claystones the knowledge already gained on clays. A new method of determining in the laboratory the thermal pressurization coefficient is described and the data obtained allow completing existing data in the field. Finally, the hollow cylinder apparatus makes it possible to demonstrate that the good self-sealing properties of clays and claystones can be extended to temperature effects, an important conclusion in terms of performance assessment.

## 1. Introduction

Clays and claystones are considered in some countries (including Belgium, France and Switzerland) as a potential host rock for high activity long lived radioactive waste disposal at great depth (Tsang et al. 2012). One of the aspects to deal with in the performance assessment (Yu et al. 2013) of high activity long lived radioactive waste disposals is related to the effects on the host rock of the temperature elevation due to the placement of exothermic wastes. In drained conditions, the volume change response of clays is known to be sensitive to temperature elevation with respect to the overconsolidation ratio, as will be commented later on. Due to the low permeability of clays and the even lower one of claystones, drained conditions only occur under very slow temperature increases. The response is then often partially drained or undrained. In such conditions, the generation of thermal pore pressure (also called thermal pressurization) is also to consider, given that the resulting decrease in effective stress may have some negative consequences in terms of stability conditions around the gallery.

In drained conditions, the question about the thermal volume change response of claystones is to know how they react and how far existing data on clays can be relevant for claystones. Due to their smaller permeability (one order of magnitude less than stiff clays, around $10^{-13}$ m/s), there is very few available data on fully saturated and drained thermal tests on claystone. To perform these difficult tests, a special hollow cylinder triaxial apparatus was developed and used, as will be described further on.

In undrained conditions, there are rather few data available concerning the thermal pressurization ratio of clays. This is even truer in claystones, also due to the difficulty of properly saturating samples because of their very small porosity.

Another important aspect in terms of performance assessment is the potential effects of temperature elevation on the crack network that characterises the Excavation Damaged Zone (EDZ) around galleries. This potential impact has been investigated in detail during the European research project TIMODAZ, some results of which will be presented here. The critical aspects concern i) the potential effect of water dilation on crack openings, keeping in mind the possibility of shear reactivation along shear cracks and ii) the effect of temperature on the self sealing properties of the clays and claystones.

This paper presents some results concerning the thermal response of clays and claystones under both drained and undrained conditions. The role of shear cracks on permeability properties at elevated temperature is also presented.

## 2. Waste disposal in clays and claystones

As an example, the underground research laboratory (URL) of ANDRA (the French Agency for Radioactive Waste Management) is located at a depth of 490 m in a layer of Callovo-Oxfordian claystone presented in Fig. 1. As seen in the Figure, the layers have regularly sedimented with a slight tilting. Other European URLs presently active in clays and claystones include the Mont-Terri URL in the Opalinus clay in Switzerland operated by the Swiss Agency (NAGRA) and the Mol URL operated by Euridice (ONDRAF-SCK) in Belgium (Tsang et al. 2012).

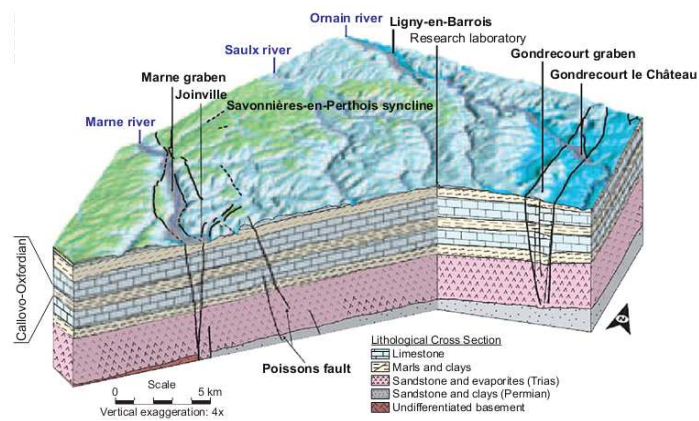

**Fig. 1.** The Callovo-Oxfordian claystone layer in the area of Bure (ANDRA, after Tsang et al. 2012).

The French principle of high activity waste repository by means of horizontal galleries is presented in Fig. 2. The design of this system is strongly conditioned by the exothermic properties of the radioactive waste and by the rule adopted in performance assessment to maintain below 100°C the temperature in the host rock in the close field. For this reason, it is not possible to excavate the galleries too close one from another. Also, some dummy canisters (with no waste inside) are placed among heat emitting canisters to lower the temperature elevation. This temperature constraint has a significant economic impact since it imposes longer galleries excavated at larger distances one from another, resulting in a larger occupation area at a given level.

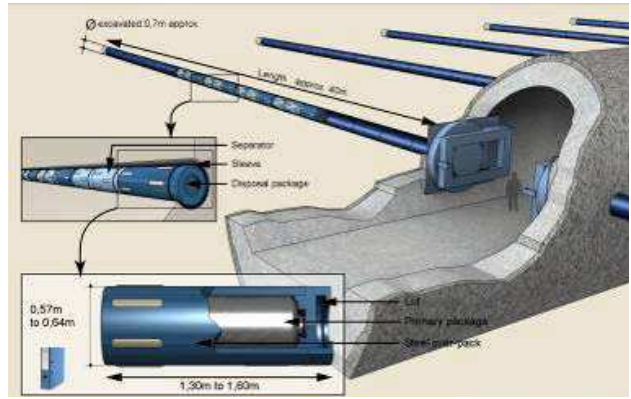

**Fig. 2.** French concept of horizontal galleries (ANDRA 2005)

The results of calculations about the thermal perturbation due to the heat emitted by high activity radioactive wastes in the French concept are presented in Fig. 3 (ANDRA 2005). Since the thermal conductivity of the claystone is well known and fairly homogeneous, thermal calculations carried out by using the finite element method are known to be reliable. As seen in the Figure, it is estimated, with the inter-gallery distance selected, that a maximum temperature of 90°C will be quickly reached after around 15 years in the close field. Temperature homogeneity starts prevailing at 40-45°C after 1 000 years and at 30-35°C after 3 000 years. The come back to the initial temperature close to 5°C is planned after 50 000 years.

Another illustration of the thermal phase is provided in Fig. 4 in which one can see that closure will be made after a period of around 30-100 years of the exploitation phase with an increasing temperature phase comprised between 100 and 300 years followed by the come back to natural conditions after 50 – 100 000 years.

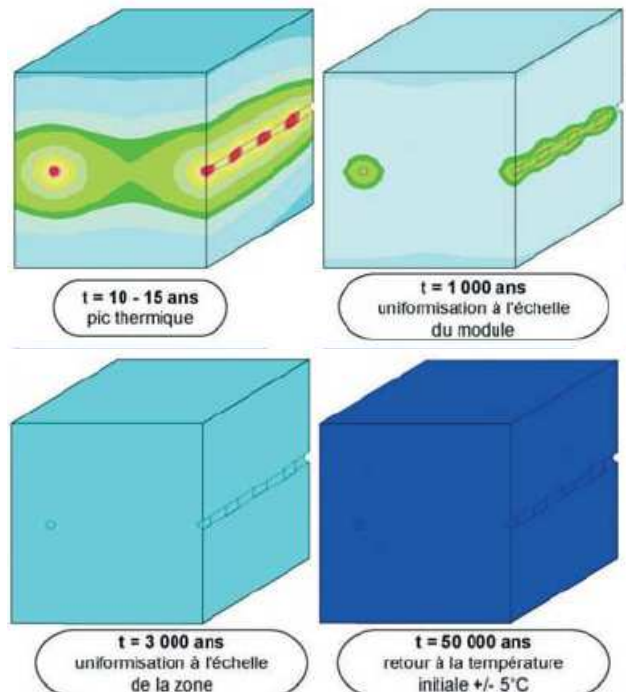

**Fig. 3.** Temperature decrease around galleries (ANDRA 2005)

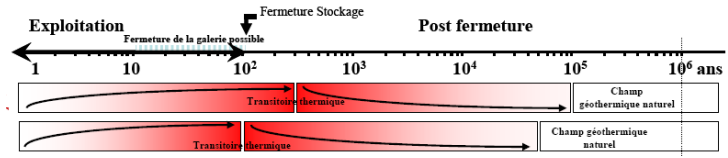

**Fig. 4.** Time schedule of thermal phenomena (ANDRA 2005)

Another important phenomenon observed during and after the gallery excavation if the damage created around the galleries in what is called the Excavation Damaged Zone.

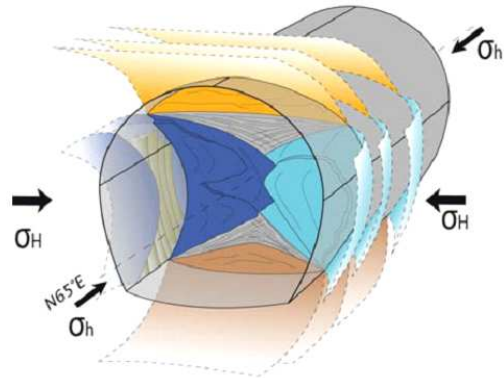

**Fig. 5.** Conceptual model of the induced fractures network around a drift parallel to the horizontal minor stress (Armand et al. 2013).

Fig. 5 shows a schematic representation of the fracture network observed around a gallery excavated parallel to the horizontal minor stress $\sigma_h$ (N 65° E), given that the in-situ stress in the Bure URL are as follows (Wileveau et al. 2010):
- Minor horizontal stress $\sigma_h$ = 12.4 MPa
- Major horizontal stress $\sigma_H$ = 12.7-14 MPa
- Vertical stress $\sigma_v$ = 12.7 MPa
- Pore pressure $u$ = 4.9 MPa

This configuration of parallel periodic inclined fractures along the gallery axis is called an "herringbone" pattern, it has also been observed in other clay host rock like Boom clay or Opalinus clay.

The extension of fractures around the same gallery is also represented in Fig. 6. The figure also makes the distinction between shear fractures with a larger extension (0.6 drift diameter on top, 0.8 drift diameter in bottom) compared to traction fractures (0.3 drift diameter).

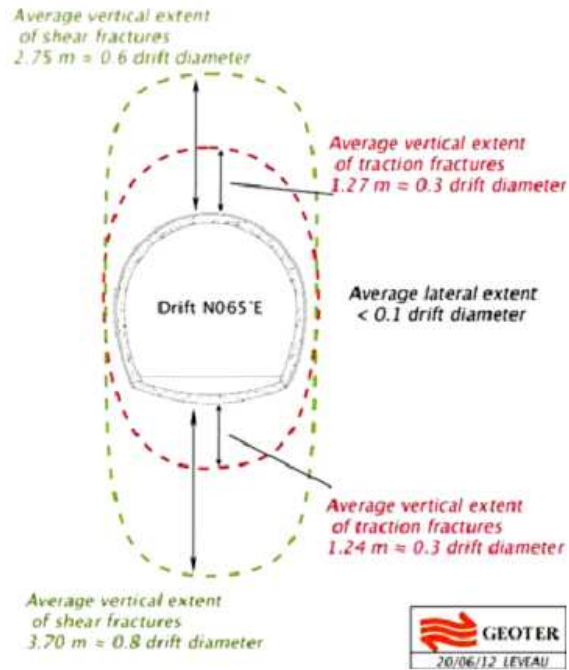

**Fig. 6.** Extent of fractures around a gallery parallel the the minor horizontal stress (Armand et al. 2013)

## 3. Thermal behaviour of clays

### 3.1. Thermal volume changes

The pioneering work of Hueckel and Baldi (1990) on Boom clay samples demonstrated that the volume changes of natural clays under constant isotropic effective stress depended on the overconsolidation ratio, independently of the applied stress. A thermal plastic contraction is observed in normally consolidated clays and an elasto (expansion) plastic (contraction) response is observed in overconsolidated clays, depending of the overconsolidation ratio.

This has been confirmed in Boom clay by Sultan et al. (2002) as shown in Fig. 7 that presents the results of drained thermal tests run on samples at various overconsolidation ratios in which the temperature elevation was very slowly applied (1°C/h) to ensure fully drained conditions.

In the overconsolidated regime, the Figure also shows that the temperature of the expansion-contraction transition increases with the overconsolidation ratio as accounted for in the thermomechanical model of Cui et al. (2000).

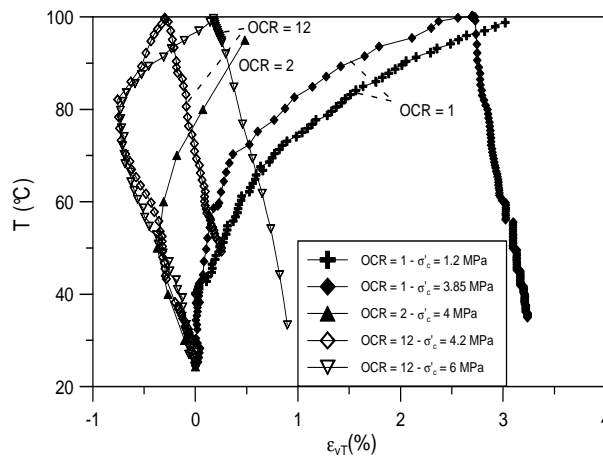

**Fig. 7.** Thermoelastoplastic response of Boom clay (Sultan et al. 2002)

Whereas the thermoelastic expansion can be easily interpreted in terms of thermal dilation of the solid phase (see thermal dilation coefficients in Table 1), the plastic contraction is more difficult to interpret. It seems that the role of water adsorbed along the clay particles plays an important role. There would be two modes of separating the adsorbed water from the (plastic) clay particles along which they are bonded: either by compression and squeezing out (effect of preconsolidation) or by heating, given that in general temperature reduces the physico-chemical interactions between molecules (consider for instance the effects of temperature on the Brownian movement and the changes in properties of gases when submitted to temperature elevation).

### 3.2. Thermal pressurization

The thermal pressurization of pore water in low porosity clays or claystones submitted to temperature elevation is a consequence of the significant difference between the thermal expansion coefficient of water ($27 \times 10^{-5}$ °C$^{-1}$) and that of the solid grains (between 1 and $3.4 \times 10^{-5}$ °C$^{-1}$), see Table 1 (Ghabezloo and Sulem 2009). The Table is completed by the values of the compressibility of each mineral that will be used further on.

**Table 1.** Thermal expansion and compressibility coefficients of some typical minerals present in clays and claystones

| Mineral | Thermal expansion coefficient (°C$^{-1}$) | Solid compressibility (GPa$^{-1}$) |
|---|---|---|
| Clay | $\alpha_s = 3.4 \times 10^{-5}$ (McTigue 1986) | $c_s = 0.02$ (Skempton 1960, McTigue 1986) |
| Quartz | $\alpha_s = 3.34 \times 10^{-5}$ (Palciauskas and Domenico, 1982) | $c_s = 0.0265$ (Bass 1995) |
| Calcite | $\alpha_s = 1.38 \times 10^{-5}$ (Fei 1995) | $c_s = 0.0136$ (Bass 1995) |
| Felspar | $\alpha_s = 1.11 \times 10^{-5}$ (Fei 1995) | $c_s = 0.0145$ (Bass 1995) |
| Water | $\alpha_w = 27 \times 10^{-5}$ (Spang 2002) | $c_w = 0.447$ (Spang 2002) |

One can see that the water expansion coefficient is almost one order of magnitude higher than that of the minerals and that the coefficient of quartz and clay are comparable and larger that that of calcite or feldspar.

Given that the structure of claystones is generally characterized by a clay matrix containing grains (of quartz, calcite and feldspar), one can suspect some significant differential expansions in the solid phase at clay-calcite interfaces (the most frequent) and also at clay-felspar interfaces.

**Table 2.** Thermal expansion coefficients of various soils and rocks

| Material | Thermal expansion coefficient $\Lambda$ (MPa/°C) | Reference |
|---|---|---|
| Clay | 0.01 | Campanella and Mitchell (1968) |
| Boom Clay | 0.06 | Vardoulakis et al. (2002) |
|  | 0.019 | Lima et al. (2010) |
| Opalinus claystone | 0.1 | Muñoz et al. (2007) |
| Sandstone | 0.05 | Campanella and Mitchell (1968) |
| Kayenta Sandstone | 0.59 | Palciauskas and Domenico (1982) |
| Rothbach sandstone | From 0.25 to 0.025 | Ghabezloo and Sulem (2009) |
| Clayey fault gouge | 0.1 | Sulem et al. (2004, 2007) |
| Intact rock at great depth | 1.5 | Lachenbruch (1980) |
| Mature fault at 7 000m depth | Intact fault wall : 0.92 Damaged fault wall : 0.31 | Rice (2006) |

Ghabezloo and Sulem (2009) gathered some values of the thermal pressurization coefficient $\Lambda = \Delta p/\Delta t$ (MPa/°C) measured in various soils and rocks and presented in Table 2. The Table has been completed by their own values on the Rothbach sandstone, by a value on Boom clay recently obtained by Lima et al. (2010) and by a value on Opalinus claystone deduced from the data of Muñoz et al. (2009).

Some large values have been obtained in rocks but the values in clays are between 0.01 and 0.1MPa/°C. Quite different values are given for clays, in particular in Boom clay, comparing the data of Vardoulakis et al. (2002) (obtained from experimental data of Sultan 1997) and that of Lima et al. (2010). Authors showed that the thermally induced pore pressure did not only depend on the mineral composition and porosity of the rock, but also on the stress state, the range of temperature variation and the previously induced damage.

The pressure dependency of the compressibility of both rock and water and the temperature dependency of the pore water compressibility appeared to play an important role, as shown by Ghabezloo and Sulem (2009) who provided values between 0.25 and 0.025MPa/°C at temperatures between 20 and 70°C for the Rothbach sandstone.

## 4. Materials

The main characteristics of the three clays and claystones considered are presented in Table 3. The data of Table 1 show a difference between on the one hand the Boom clay (BC), a stiff clay with significantly higher porosity and smaller mechanical parameters and on the other hand with the Opalinus clay (Opa) and the Callovo-Oxfordian (COx) claystones of smaller porosity, water content and permeability with higher calcite content and mechanical parameters.

Whereas the Boom clay is from the Rupelian age (Oligocene epoch) and is around 30 millions years old, the Callovo-Oxfordian and Opalinus claystones are significantly older (162 and 173 millions years respectively, Gens 2012) and are from the Callovo-Oxfordian and Aalenian ages (Jurassic epoch), respectively.

**Table 3.** Characteristics of clays and claystones
(Gens et al. 2007)

|  | Opalinus clay | Callovo-Oxfordian claystone | Boom clay |
|---|---|---|---|
| Dry density (Mg/m$^3$) | 2.22-2.33 | 2.21-2.33 | 1.61-1.78 |
| Calcite content | 6-22% | 23-42% | 0-3% |
| Porosity | 13.5-17.9 | < 13 | > 30 |
| Water content | 4.2-8% | < 5.5% | > 9.5% |
| Young's modulus, MPa | 4000 - 10000 | 4000-5600 | 200-400 |
| UCS, MPa | 4-22 | 20-30 | 2 |
| Permeability, m/s | 1-5 × 10$^{-13}$ | 1-5 × 10$^{-13}$ | 2-5 × 10$^{-12}$ |

The difference in mechanical properties between the BC clay and the claystones (COx and Opa) is due to the much longer diagenesis period claystones have been submitted to. Another significant difference is related to the higher carbonate content that they contain (highest in COx) and that also results in stronger mechanical properties. The carbonates acting as a bonding agent are carbonates reprecipitated from their initial granular state. The COx claystone also contains a significant proportion of qualrtz grains (around 20%).

Due to smaller porosities, claystones also have permeability smaller by one order of magnitude compared to the BC clay. As commented before, these very low porosities and permeability values makes quite difficult fully saturated and drained conditions.

## 5. Test in drained conditions

### 5.1. Hollow cylinder triaxial apparatus

The hollow cylinder triaxial apparatus specially designed for very low permeability geomaterials was developed within the framework of the TIMODAZ European project (Monfared et al. 2011). In this apparatus, the same confining pressure is applied in the inner and outer lateral faces of the cylinder, providing standard triaxial stress conditions.

The principle of the apparatus is shown in the scheme of Fig. 8 in which lateral drainage is ensured by geotextiles placed on the inner and outer faces of the cylinder.

Heating was made by using an electric coil placed around the metal cell (not represented in the Figure) and by thermally isolating the cell from the ambient temperature by using an isolating cover. Temperature control (± 0.5°C) was carried out based on a temperature measurement gauge located inside the cell. Volume changes were followed from local displacement measurements by means of LVDT sensors along the external face of the cylinder, as shown in Fig. 9 that also shows the radial hydraulic connections used to ensure enhanced lateral drainage.

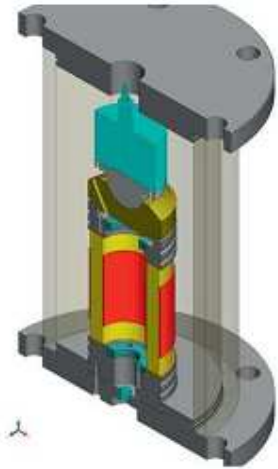

**Fig. 8.** Hollow cylinder triaxial cell (Monfared et al. 2011)

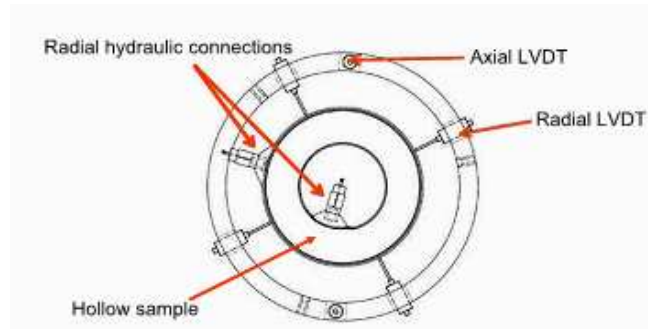

**Fig. 9.** Local strain measurements by using LVDTs

### 5.2. Saturation procedure

An important preliminary condition to fulfil before running drained or undrained isotropic or triaxial compression tests is to ensure full saturation of the sample. In stiff clays like the Boom clay, this condition is generally controlled by determining the Skempton $B$ coefficient by monitoring the increase in pore pressure $\Delta u$ resulting from an increase in applied confining pressure $\Delta\sigma$ according to the relation:

$$B = \Delta u/\Delta\sigma \qquad (1)$$

In soils, $B$ values above 95% are considered satisfactory (see for instance Sultan et al. 2010 on Boom clay).

The determination of $B$ values does not seem to be common practice in claystone testing in which the saturation conditions are apparently not fully known, like also the drainage conditions. Indeed, claystone samples appear to be provided in the laboratory in unsaturated conditions, due to the combined and successive effects of air coring, transport, storage and trimming. As an example, Mohajerani et al. (2012) determined initial values of degree of saturation around 70% in the Cox claystone.

Indeed, claystones are very sensitive to changes in water content. Like unsaturated soils, they become significantly stronger when desaturated. As shown in Fig. 10, Pham et al. (2007) observed increases in Unconfined Compression Shear (UCS) from 28 to 58 MPa in samples submitted to desaturation by imposing controlled humidities RH of 98 and 32% respectively, with corresponding water contents decreasing from 5.24 to 1.65%.

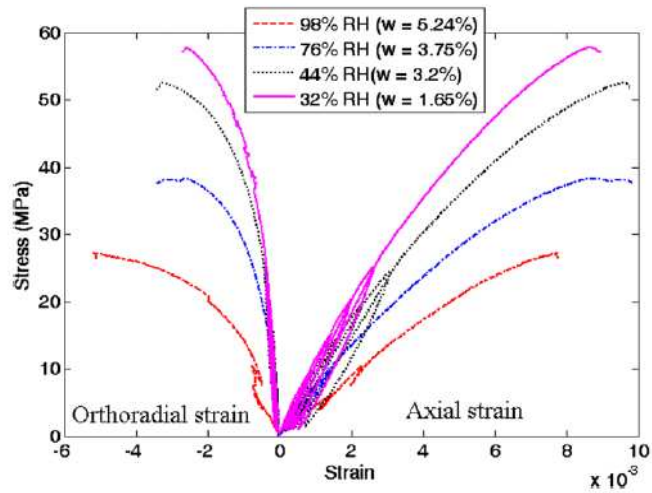

**Fig. 10.** Water sensitivity of COx clay, UCS tests (Pham et al. 2007).

The data of Fig. 10 clearly indicate that partial saturation may overestimate the mechanical characteristics of claystones. Due to very low permeability, claystone are really difficult to fully saturate. Homand (2008) quoted a period of time of one year to monitor a pressure response on the top of a 76 mm high triaxial sample submitted to a pressure at its base. To cope with the difficulty of satisfactorily saturating claystone specimens within a reasonable period of time, the option chosen was to reduce the drainage length of the specimen. In standard triaxial testing, this drainage length is equal to half the height of standard specimens (76 mm) i.e. 38 mm. The reduction in drainage length was obtained by using hollow cylinder specimens drained on top, bottom and also along the inner and outer faces of the hollow cylinder. The sample has internal and external diameters equal to 60 and 100 mm respectively, giving a drainage length of half the cylinder thickness, i.e. 10 mm.

Given that the phenomenon is governed by Richards' equation and is of a diffusion nature, a first estimation of the reduction in time can be obtained by considering that the reduction by 4 of the drainage length turned into a reduction by 16 of the time to dissipate. In such conditions, the 1 year period of time mentioned above reduces in around three weeks, a much more reasonable period. More detailed calculations will be presented later on.

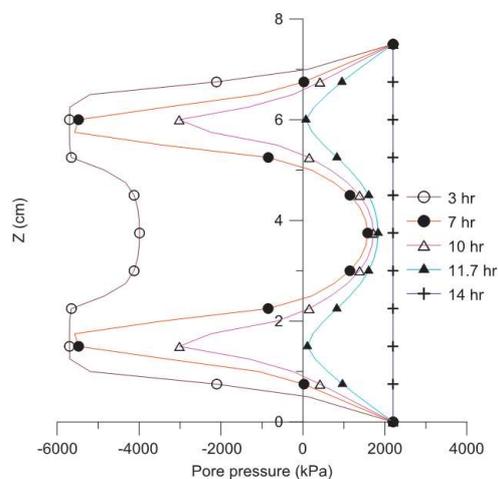

**Fig. 11.** Computed pore pressure profiles during saturation (Monfared et al. (2012)

Monfared et al. (2011) provided a numerical simulation (resolution of Richards's equation by using the finite differences method) of the saturation process of a sample of Opalinus clay from an initial state characterised by a suction of 5.7 MPa (initial degree of saturation of 83%) when saturation is made up by applying a 2.2 MPa back pressure on top, bottom and through the lateral inner and outer drainages. Calculations show that the back pressure imposed on top, bottom and along the inner and outer faces of the sample is finally homogeneously imposed in the whole sample after 14 h, quite a short period showing the efficiency of the reduced drainage length and of the enhanced drainage.

As commented above, a global checking of satisfactory saturation can be obtained by determining the Skempton coefficient $B$. In the framework of poroelasticity (see for instance Detournay and Cheng 1993), the expression of $B$ is as follows:

$$B = \frac{\Delta u}{\Delta \sigma} = \frac{c_d - c_s}{c_d - c_s + \phi(c_w - c_s)} = \frac{c_d - c_s}{S} \tag{2}$$

in which S is the storage coefficient:

$$S = c_d - c_s + \phi(c_w - c_s) \tag{3}$$

and $\phi$ the porosity, $c_d$ the drained compressibility of the claystone, $c_s$ the compressibility of the solid grains and $c_w$ that of water. Parameter $c_s$ can be determined by a so-called unjacketed compression test in which the back pressure and the confining stress are simultaneously and equally increased while measuring the volume changes of the specimen.

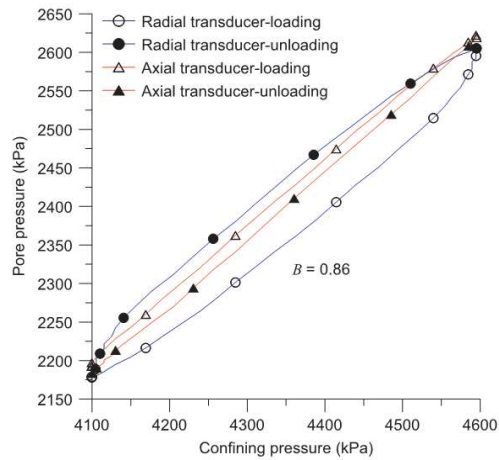

**Fig. 12.** Undrained isotropic compression test of the Opalinus clay: determination of the Skempton coefficient (Monfared et al. 2011)

Fig. 12 presents the results of an isotropic "undrained" compression test in which the pore pressure increment $\Delta u$ is presented with respect to the increment in total stress $\Delta \sigma$. In soils, good saturation corresponds to $B$ values close to one. In porous stones, this is no longer true given the form of expression (2) and the fact that in porous stones $c_d$ is no longer negligible compared to $c_s$.

Actually, an "undrained" test is in fact a test in which the drainage valves connected to the cell base are closed. It is not fully undrained given that some water exchanges occur between the porosity of the compressed specimen and that of the porous elements surrounding it. This effect has also to be accounted for when comparing the experimental value and the theoretical expression given in (2). Detailed calculations made in Monfared et al. (2011) show that a corrected theoretical value of 0.89 is obtained that compares favourably with the slope ($B = 0.86$) calculated from the data of Fig. 12 (with $c_w = 0.449$ GPa$^{-1}$, $c_d = 0.348$ GPa$^{-1}$, $c_s = 0.02$ GPa$^{-1}$ for the Opalinus clay).

Note that satisfactory conditions in swelling soils can only be ensured under in-situ stresses to avoid any swelling (see for instance Delage et al. 2007 in the Boom clay). In claystones, it is well known that swelling can induce significant damage. The reduction in time of the saturation process then becomes an important issue given that it reduces the time of immobilisation of the triaxial device.

**5.3. Drained conditions**

Satisfactory drainage conditions also have to be ensured. Calculations of pore pressure dissipation during an isotropic compression test run at a rate of 0.5 kPa/mn were also carried out by Monfared et al. (2012).

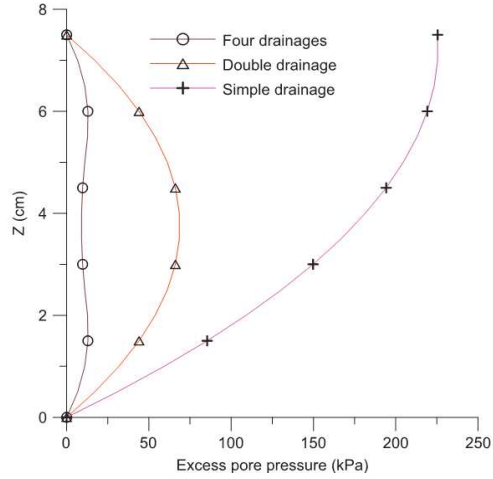

**Fig. 13.** Computed excess pore pressure profiles in half the thickness of the sample during isotropic compression under different drainage conditions (Monfared et al. (2012)).

Fig. 13 shows the excess pore pressure calculated once a 500 kPa isotropic stress is applied in a sample having a permeability typical of claystones equal to $2.5 \times 10^{-20}$ m$^2$.

Note that pore pressure dissipation calculations in claystones are different than that conducted in clays given that water and the solid phases cannot be considered as infinitely rigid compared to the drained compressibility of the sample. In the framework of isotropic poroelasticity, the pore pressure changes during isotropic compression are given by:

$$\frac{du}{dt} = B\frac{d\sigma}{dt} + \frac{k_s}{\mu_w S}\nabla^2 u \tag{4}$$

in which $B$ is the Skempton coefficient (2), $k_s$ the saturated permeability, $\mu_w$ the water viscosity and $S$ the storage coefficient (3).

The data of Fig. 13 show the excellent efficiency of the enhanced drainage system. Whereas a 70 kPa excess pore pressure is calculated with the standard configuration (drainages on top and bottom with a height of 70 mm and a drainage length of 35 mm), the lateral drainages (drainage length of 10 mm) reduce the excess pore pressure to less than 20 kPa. This shows that i) satisfactory drainage can be ensured in low permeability claystones and ii) the pore pressure field remains homogeneous during the isotropic compression test.

## 6. Test in undrained conditions

To investigate thermal pressurization, an isotropic thermal compression device simpler than the hollow cylinder was developed. The aim of the test is to measure the thermal pore pressure developed when the specimen is submitted to temperature elevation under a given stress level. This thermal pore pressure generation is due to the fact that the thermal expansion coefficient of water is significantly higher than that of the solid phase.

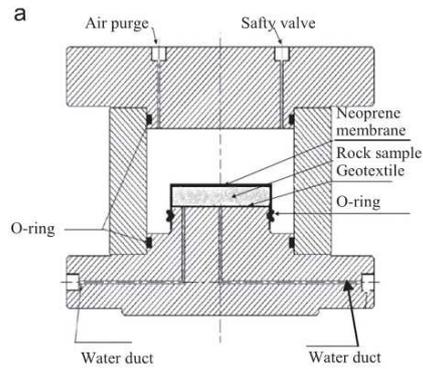

**Fig. 14.** Isotropic compression cell (Mohajerani et al. 2012)

The cell is presented in Fig. 14 and the heating and loading device in Fig. 15. The cell is a standard high pressure isotropic compression cell accommodating a disk shaped specimen (oedometer type) having a diameter of 70 mm and a height of 10 mm. Drainage is ensured by a geotextile at the base of the sample resulting in a small length of drainage of 10 mm. Two water ducts connect the sample bottom to either a pressure volume controller for back pressure (PVC, GDS brand) or a pressure gauge.

Fig. 15 presents the connections to the two PVC and the temperature controlled bath (± 0.1°C) in which the cell is immersed. Note that the pressure transducer is located outside the bath.

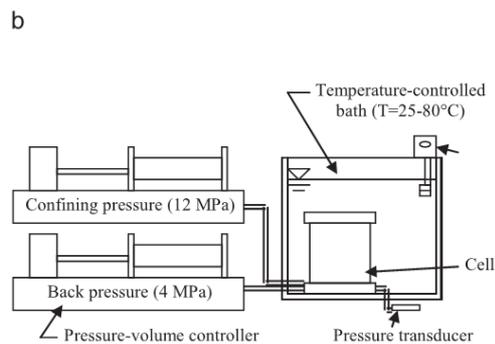

**Fig. 15.** Stress and temperature control (Mohajerani et al. 2012)

Like in the hollow cylinder device, special care has been devoted to the saturation procedure that was carried out under stress conditions close to in-situ. Given that the drainage length is equal, similar saturation period were observed compared to the hollow cylinder device. The determination of satisfactory values of the Skempton $B$ parameter was also made.

## 7. Thermal volume changes in claystones

To complete the few available literature data about fully saturated and drained thermal tests in low permeability claystones, a program including drained heating tests under in-situ stress conditions was carried out in the hollow cylinder apparatus. The question as to whether claystone behave as clays was investigated.

### 7.1. Opalinus clay

The response of a specimen of Opalinus clay to a drained thermal cycle between 25 and 80°C under in-situ mean stress conditions (total mean stress $p$ = 4.1 MPa and pore pressure $p_w$ = 2.2 MPa) is presented in Fig. 16. The test was conducted at a heating rate of 1°C/h so as to ensure fully drained conditions with the enhanced drainage suystem (see above).

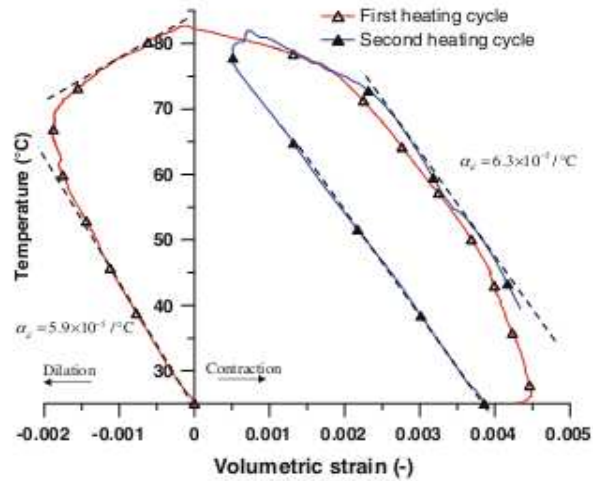

**Fig. 16.** Drained thermal volume changes of Opalinus clay under in-situ mean stress conditions (Monfared et al. 2012)

The Figure shows that the first temperature elevation phase is characterized by a linear thermo-elastic expansion that is typical of overconsolidated samples between 25 and 65°C, followed by a plastic contraction between 65°C and the maximum temperature applied, i.e. 83°C. The subsequent cooling phase is characterized by a thermo-elastic contraction with a slope approximately parallel to that of the first heating phase, confirming the reversible nature of the response. The corresponding thermal expansion coefficients are $\alpha_d = 5.9 \times 10^{-5}$ °C$^{-1}$ along the heating phase and $6.3 \times 10^{-5}$ °C$^{-1}$ along the cooling phase. Note that these thermal expansion coefficient are comparable, although somewhat larger, to that of the solid phases present in the claystone (between 1 and $3.4 \times 10^{-5}$ °C$^{-1}$, see Table 1).

The thermal irreversible contraction observed during the first heating phase is comparable to that observed in normally consolidated plastic clays (e.g. Baldi et al. 1988 and Delage et al. 2000 on Boom clay). In fine grained soils, this phenomenon is understood as a thermal consolidation of the sample which corresponds to the rearrangement of the grains after a critical temperature. It seems that the claystone has kept the memory of its maximum supported temperature like overconsolidated soils conserve the memory of the maximum supported load. The sample once heated up to 83°C after the first cycle, keeps expanding up to 80°C (the new maximum temperature supported) during the second cycle, forgetting the previous temperature threshold of 65°C.

It is interesting to relate this first temperature threshold to the maximum temperature previously experienced by Opalinus clay. From geological arguments the maximum burial depth of the Opalinus Clay at Mont Terri is 1350 m (Tim Vietor, personal communication). Assuming a geothermal gradient of about 0.03°C/m, the value of 65°C appears as a plausible maximum temperature experienced by the material before our test. The observed behaviour is typical of thermal hardening, with an elastic thermal expansion observed below the maximum supported temperature, followed by a plastic contraction at yielding once the maximum temperature is attained. Note that the concept of thermal yield has been described in the model developed by Cui et al. (2000) for normally and overconsolidated soils through the mobilization of a specific TY yield curve coupled to a loading yield curve similar to that developed by Hueckel and Borsetto (1990). This approach could certainly be adapted to claystone by accounting for the strong post-sedimentation diagenetic links that characterizes claystones.

The elasto-plastic dilation coefficient ($\alpha_d^{ep}$) of the sample can be estimated by considering the volumetric response of the sample between 68°C and 83°C during the first heating phase which gives a value of $-1 \times 10^{-4}$ /°C. Thus the plastic thermal expansion coefficient of Opalinus claystone is obtained from (Sulem et al. 2007):

$$\alpha_d^p = \alpha_d^{ep} - \alpha_d \qquad (5)$$

which gives a value of $-1.64 \times 10^{-4}$ °C$^{-1}$. The obtained plastic thermal dilation coefficient for Opalinus claystone is smaller than that for Boom clay ($-7 \times 10^{-4}$ °C$^{-1}$ after Delage et al. 2000) and clayey gouge from Aigion fault ($-3.8 \times 10^{-4}$ °C$^{-1}$ after Sulem et al. 2007) which confirms again the important role of post-sedimentation diagenetic bonds in claystones compared to plastic clays.

### 7.2. Callovo-Oxfordian claystone

The measurements obtained from the radial and axial LVDTs during a drained heating tests on a COx claystone specimen (heating rate 0.5°C/mn) submitted in the hollow cylinder apparatus to stress conditions close to in-situ ($\sigma_v$ = 12.7 MPa, $\sigma_H$ = 12.7-14 MPa, $\sigma_h$ = 12.4 MPa, $u$ = 4.9 MPa) are presented in Fig. 17 (Mohajerani et al. 2013). This condition was approximately fulfilled by imposing a 12 MPa confining stress and a 4 MPa pore pressure. As seen in the Figure, the monitoring of the LVDTs was interrupted between 54°C and 68°C due to electric shortage. They worked again between 68°C and 80°C.

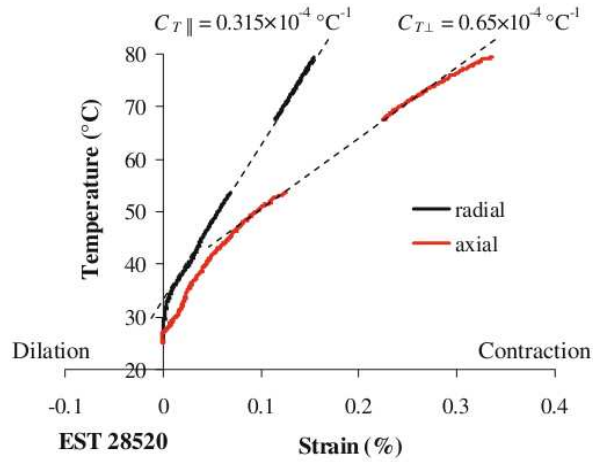

**Fig. 17.** LVDT measurements, drained heating test on the COx claystone under in-situ stress conditions (Mohajerani et al. 2013).

Both LVDT measurements show that a contraction of the COx claystone occurs from the beginning of the test, with axial strains (perpendicular to bedding) larger than radial strains (parallel to bedding) showing some degree of anisotropy in the thermal response. At 80°C, the axial strain is equal to 0.33% compared to 0.16% for the radial strain. Keeping in mind that the Callovo-Oxfordian claystone is made up of grains embedded in a clay matrix, this shows that the direction perpendicular to bedding (characterised by a slope $C_{Tperp}$ = 0.65×10$^{-4}$ °C$^{-1}$) is more sensitive to thermal changes than that parallel to bedding ($C_{Tparr}$ = 0.315×10$^{-4}$ °C$^{-1}$).

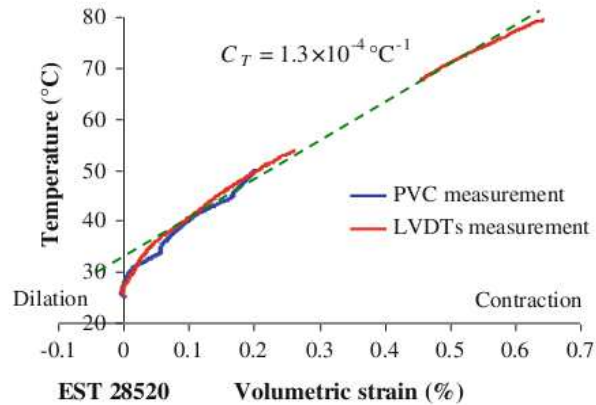

**Fig. 18.** Drained heating test on the Cox claystone (Mohajerani et al. 2013)

Fig. 18 shows the sample volume changes calculated from the LVDT data of Fig.4 and from the PVC water exchanges (only up to 50°C because a leak occurred at higher temperature). The thermal contracting plastic behaviour is confirmed with a good correspondence between the two measurements up to 50°C. An average value of $1.3\times10^{-4}\,°C^{-1}$ was found for parameter $C_T$. The LVDTs response indicates a total volumetric strain of 0.64% at 80°C.

Fig. 19 presents the responses in axial and radial strains from LVDTs of a drained heating test under in-situ stress conditions of a COx sample previously submitted to a loading cycle (12 – 18 – 12 MPa). In this case, the thermal volume changes are characterised by an initial expansion between 25 and 32°C followed by a contraction. The dilation perpendicular to bedding is almost 4 times larger than that parallel to bedding, with respective linear dilation coefficients $\alpha_\perp=0.395\times10^{-4}\,°C^{-1}$ and $\alpha_{//}=0.111\times10^{-4}\,°C^{-1}$.

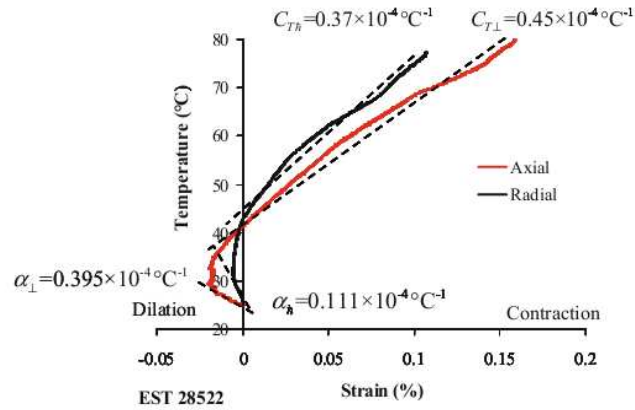

**Fig. 19.** LVDT measurements, drained heating test on a COx sample previously submitted to a (12-18-12 MPa) loading cycle.

The transition from expansion to contraction occurs at 33°C. Afterwards, contraction appears to be slightly larger perpendicular to bedding ($C_{T\perp}=0.45\times10^{-4}\,°C^{-1}$) than parallel to bedding ($C_{T//} = 0.37\times10^{-4}\,°C^{-1}$) with an anisotropy ratio significantly smaller than during contraction. This reveals smaller anisotropy effects compared to the data of Fig. 17.

The volume changes derived from the LVDTs and from the PVC are compared in Fig. 20. They show similar trends, with however smaller dilation from PVC data. The thermal expansion coefficient ($\alpha$) calculated during the expansion phase is about $0.6\times10^{-4}\,°C^{-1}$ (PVC and LVDTs) and the elastoplastic thermal contraction coefficient ($C_{T1}$) is $1\times10^{-4}\,°C^{-1}$ from the

The data of test EST28520 in Fig. 18 show that, like normally consolidated clays, the COx claystone exhibits a plastic contraction when heated in drained conditions between 25 and 80°C under in-situ stress conditions. Actually, this trend is not in agreement with the in-situ measurements of thermally induced pore pressure that appeared to be correctly modelled under a thermoelastic (dilating) hypothesis (Gens et al. 2007, Armand 2012). If confirmed, this thermal contraction should indeed have significant effects on the stress redistribution in heated zones far enough from the waste disposal galleries and remaining in a stress state close to the initial one. A possible reason of this difference between in-situ and laboratory data could be related to the quality of extracted samples in general, given that claystones are known to be very sensitive to damage induced by stress release and drying during coring. As seen above, significant precautions have been taken here to resaturate the laboratory samples under in-situ stress conditions s as to minimise any damage due to hydration and swelling. Indeed, further investigation is necessary in this regard.

LVDTs and $C_{T2} = 0.84\times10^{-4}\,°C^{-1}$ from the pore pressure PVC. A volumetric strain of 0.4% is reached at 80°C.

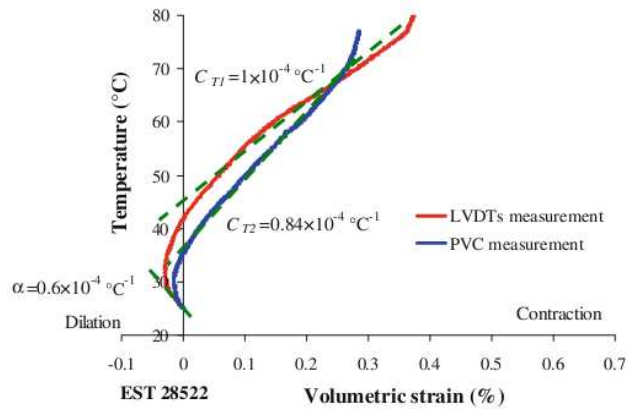

**Fig. 20.** Volume changes, drained heating test on a COx sample previously submitted to a (12-18-12 MPa) loading cycle.

Sample EST28522 was submitted to a cyclic isotropic drained loading (from 12 to 18 and then to 12 MPa) prior to drained heating, providing a drained compressibility value $c_d = 0.45\text{GPa}^{-1}$. Further drained heating between 25 and 80°C exhibited a slight initial dilating thermo-elastic response between 25 and 32°C prior to a plastic contracting phase (Fig. 20). Interestingly, this response is comparable to that of overconsolidated clays (that have also been submitted, during their geological history, to stress levels higher than that under which heating is performed). An extrapolation of this trend (to be experimentally checked) could suggest that the response of areas closer to the galleries in which stress has been somewhat released could be dilating and thermo-elastic. Again, the necessity of further thermomechanical tests to confirm this trend is mentioned, as commented above.

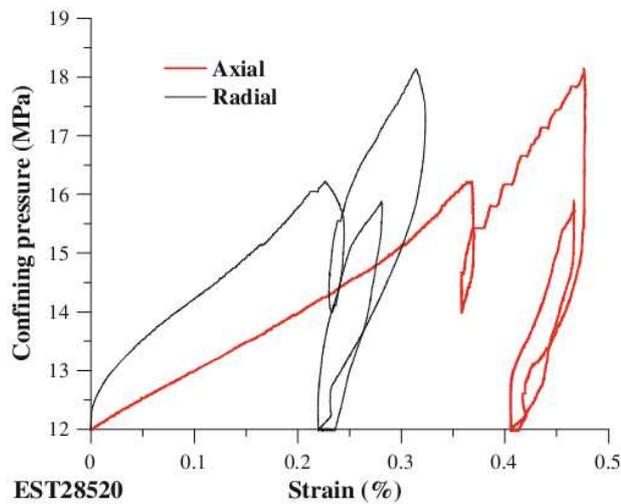

**Fig. 21.** Drained compression test at 80°C, Cox claystone.

To further investigate the effects of temperature on the volume change behaviour of claystones, a drained isotropic compression test was run at 80°C after the drained heating phase that presented a contracting plastic response (Fig. 18).

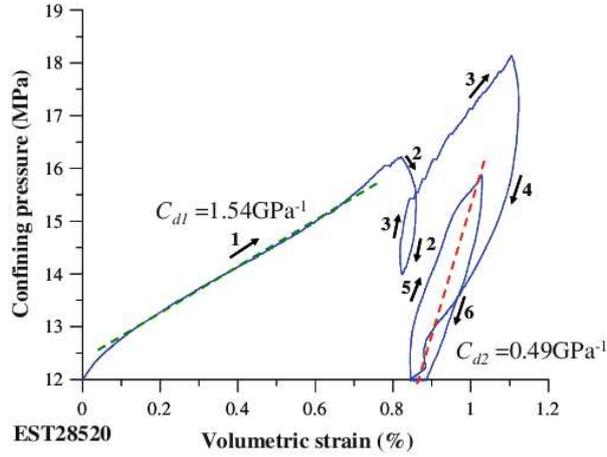

**Fig. 22.** Volume changes, drained heating test on a COx sample previously submitted to a (12-18-12 MPa) loading cycle.

The response presented in Fig. 21 again shows a significant effect of anisotropy with axial strains significantly larger than radial ones, like at ambient temperature.

The global volume changes presented in Fig. 22 show that during the first loading phase the sample exhibits higher plastic compressibility than at 25°C (1.54 GPa$^{-1}$ compared to 0.45 GPa$^{-1}$) while the elastic modulus obtained in the final cycle was about 0.49 GPa$^{-1}$, a value comparable with the elastic modulus of 0.45 GPa$^{-1}$ determined at 25°C. Once checked, this new trend should be introduced in the constitutive models used in the numerical predictions of radioactive waste disposal since this smaller compressibility at higher temperature should result in some stress redistribution in the zones that are compressed during heating.

## 8. Thermal pressurization in claystones

In a perfect undrained THM test carried out in an elastic porous material, the pore pressure increase is given by the following expression:

$$\Delta u = B \, \Delta\sigma + \Lambda \, \Delta T \tag{6}$$

where the Skempton coefficient $B$ is given by Eq. (2) and the thermal pressurization coefficient $\Lambda$ is defined by the following equation:

$$\Lambda = \frac{\Delta u}{\Delta T} = \frac{\phi(\alpha_w - \alpha_s)}{c_d - c_s + \phi(c_w - c_s)} \tag{7}$$

in which $\phi$ is the porosity, $\alpha_i$ are the thermal dilation coefficients and $c_i$ the compressibility coefficients (see equations above for complete parameter definitions).

As in the case of the determination of the Skempton $B$ coefficient, corrections are to be made in the determination of $\Lambda$ with respect to perturbations due to the effects of the porous elements in contact with the specimen (see Ghabezloo and Sulem 1999 and Mohajerani et al. 2011).

The data of Fig. 23 show the response in pore pressure obtained in a COx sample along a ramp of stepwise progressive temperature elevation between 25 and 70°C. The comparison with the response obtained with a dummy metal sample indicates that the first peak of temperature corresponds in fact to the instantaneous response of the water contained in the permeable porous elements. Some time has to be waited for to get equilibration between the pore water and that contained in the porous elements, corresponding to the drainage of pore water from the specimen to the porous elements. When considering the equilibration points, one obtains the curve presented in Fig. 24. One can observe from the results of two tests that the thermal coefficient $\Lambda$ decreases with temperature from 0.116 to 0.063 MPa/°C between 42 and 56°C in test N°1 and from 0.145 to 0.107 MPa/°C between 32 and 62°C in test N°2.

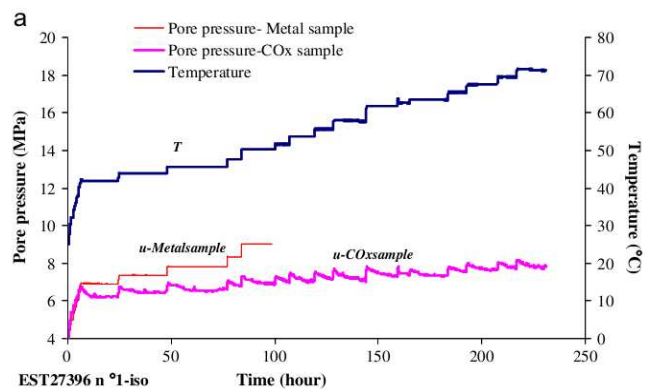

**Fig. 23.** Comparison between thermally induced pore pressure with a COx claystone and a dummy metal sample (Mohajerani et al. 2011)

As commented in more details in Mohajerani et al. (2011), the use of Eq. 7 to correctly predict the experimental changes of $\Lambda$ with respect to temperature observed in Fig. 24 requires the adoption of a non constant value of the drained compressibility $c_d$ changing with either the applied stress or temperature (see Fig. 22).

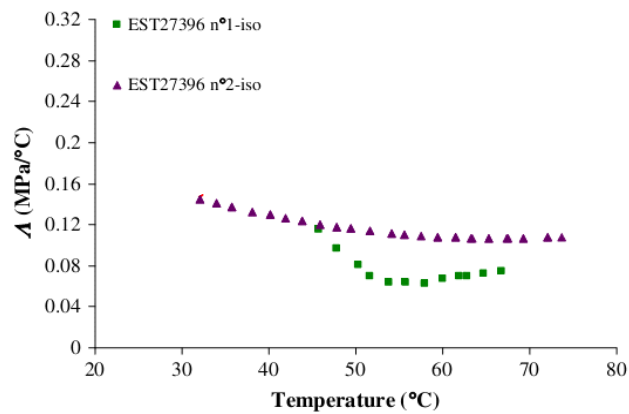

**Fig. 24.** Changes in thermal pressurization coefficient with temperature (Mohajerani et al. 2011).

Indeed, thermal pressurization (in undrained conditions) is a complex combination of increase in pore pressure that result in a corresponding decrease in effective stress that mobilises the specimen's compressibility besides the thermal dilation coefficients of water and of the solid phases.

## 9. Effect of discontinuities – self sealing/ temperature effects

One of the main interests of the hollow cylinder test is that it allows to perform permeability tests that do account for the shear plane network that affect a sheared sample. The system was then applied to the investigation of the combined effects of shear discontinuities and temperature on the permeability of a Boom clay specimen.

To do so, a transient method was adopted, based on the method proposed by Ghabezloo et al. (2009) that first consists in inducing an excess pore pressure in the sample by performing an isotropic undrained compression test. The excess pore pressure is afterwards released on one side of the sample while monitoring the pore pressure change on the other side. The permeability is evaluated through a back analysis of the experimental results by solving a diffusion problem (see Monfared et al. 2012 for more details).

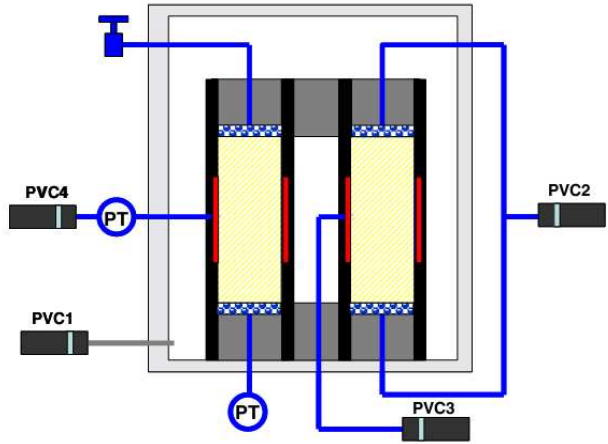

**Fig. 25.** Setting of the connection during the permeability test (Monfared et al. 2012).

Starting from the in-situ mean effective stress characterising the Boom clay layer in the Mol URL where it has been extracted ($\sigma_3$ = 3.25 MPa, $u$ = 1 MPa), the permeability tests were carried out by increasing the confining pressure from 3.25 MPa to 3.75 MPa in undrained condition. The excess pore pressure is afterwards allowed to dissipate by opening the connection of the lateral internal drainage to the PVC imposing the constant 1 MPa pressure while monitoring the pore pressure dissipation by two pressure transducers at the bottom and at the external wall of the sample (see Fig. 25). The water expelled from the sample is measured by the PVC connected to lateral internal drainage.

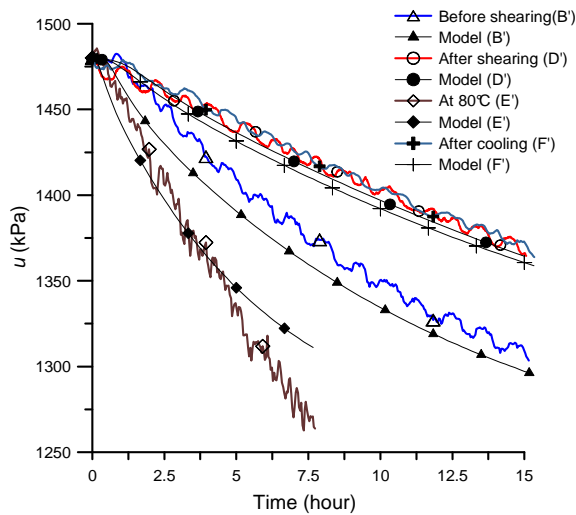

**Fig. 26.** Experimental data and model results for pore pressure dissipation, external wall transducer (Monfared et al. 2012).

The curves of pore pressure dissipation obtained in various conditions are presented in Fig. 26:
- after sample resaturation under in-situ stress
- after shearing
- after heating the sheared sample
- after cooling.

Since pore pressure measurements have been conducted both on the lateral side and at the bottom of the sample (see Fig. 25), calculations have been made with a anisotropic permeability model, providing the radial and axial permeabilities. As seen in Fig. 26, model calculations compare well with experimental data at each stage.

The slower diffusion observed after shearing compared to initial state confirms the excellent self-sealing properties of the Boom clay. As seen in Table 4, the slight decrease observed in intrinsic permeability is due to a decrease in porosity that results from the increase in mean effective pressure occurring during shearing. The faster diffusion observed at elevated temperature in the sheared sample is due to the decrease in viscosity of water with temperature. Indeed, the data of Table 4 show that the intrinsic permeability is not significantly affected

Table 4. Changes in permeability, intact and sheared Boom clay sample

|  | $k_r$ (m$^2$) | $k_z$ (m$^2$) | $\Delta\phi$ (%) |
|---|---|---|---|
| Before shearing, 25° | $1 \times 10^{-19}$ | $3 \times 10^{-19}$ | - |
| After shearing, 25° | $0.5 \times 10^{-19}$ | $2 \times 10^{-19}$ | - 0.238 |
| At 80°C | $0.8 \times 10^{-19}$ | $2.8 \times 10^{-19}$ | 0.259 |
| After cooling, 25° | $0.5 \times 10^{-19}$ | $2 \times 10^{-19}$ | - 0.259 |

These data, along with other data obtained in the Opalinus clay, show that the self-sealing properties of both the Boom clay and the Opalinus claystone are not affected by temperature elevation. This important result obtained within the TIMODAZ project is quite important in terms of performance assessment (Yu et al. 2013).

## 10. Conclusions

Various new specific experimental devices have been developed within the framework of the TIMODAZ European research project to investigate the impacts of temperature elevation in the Excavation Damaged Zone around the galleries containing high activity long lived radioactive wastes.

In some way, the option followed during this research was to extend to porous clayey rocks some well known concepts from soil mechanics that apparently had not yet been fully integrated. This approach comprised the necessity of properly saturate specimens under in-situ stress prior to testing and to ensure satisfactory drainage conditions, a challenge in low permeability. This was obtained by developing new devices characterized by a short drainage length, in particular a new hollow cylinder thermal triaxial apparatus. A short drainage length isotropic cell was also used to investigate thermal pressurization. The experimental investigation was conducted on three clays and claystones considered as potential host rocks: the Boom clay (BC) from the Mol URL (Belgium), the Opalinus clay (Opa) from the Mt-Terri URL (Switzerland) and the Callovo-Oxfordian (COx) claystone from the Bure URL (France).

A first series of tests allowed by the hollow cylinder apparatus are fully drained thermomechanical tests that allowed complete the understanding of the thermal volume changes in clays and claystones. The same principles appear to apply to claystones that should not be considered as overconsolidated clays. In the COx claystone, a contracting behaviour has been observed when heating the sample under in-situ stress, comparable to the response of normally consolidated clays. Once preloaded by a stress cycle, the COx clay exhibits a thermo-elasto-plastic response with initial thermo-elastic expansion followed by contraction at higher temperature. The Opalinus clay also exhibited a thermo-elasto-plastic response with a thermal dilation up to the maximum temperature supported by the layer during geological history, followed by plastic contraction. Thermal hardening was also evidenced during a subsequent temperature cycle.

Thermal pressurization was investigated in details, showing a decrease in thermal pressurization coefficient $\Lambda$ with increased temperature. Undrained heating tests indeed appeared as quite complex with the combined effect of thermal pore pressure build-up with the corresponding effective stress decrease that mobilises some changes in compression-swelling coefficients with both temperature and stress changes. Good understanding of thermal pressurization is necessary to better assess the stability of galleries during the thermal phase.

Both temperature and shear discontinuities do not affect the confining performance of the clays and claystones. This was expected in a stiff plastic clay like the Boom clay. In claystones, this is linked to the smectite fraction that appears to be mobilised along the shear discontinuities. The well known self-sealing performances of clays and claystones are extended to temperature effects.

Globally, results about the response of the clays and claystones have been positive in terms of performance assessment, as detailed in Yu et al. (2013). This is one of the main and successful results from the TIMODAZ project.

**Acknowledgements**

The researches presented in this paper have been obtained during the PhD works of Dr. M. Mohajerani and Dr. M. Monfared who were also supervised by Dr. J. Sulem and Dr. A.M. Tang. Dr. Monfared's PhD was supported jointly by the European TIMODAZ project (F16W-CT-2007-036449) managed by Dr. F. Bernier and Dr. X.L. Li from Euridice (Belgium) and by Ecole des ponts ParisTech. Dr. Mohajerani's PhD was jointly supported by ANDRA and Ecole des ponts ParisTech. The author also thanks Dr. T. Vietor from NAGRA, Dr. J.D. Barnichon and B. Gatmiri from ANDRA, Dr. Ghabezloo and Dr. Cui from Ecole des ponts ParisTech for stimulating discussions. The technical support provided by the technical team of the Navier/CERMES laboratory under the guidance of M. E. De Laure is also gratefully acknowledged.